\documentclass[twocolumn,final]{svjour3}

\usepackage{amssymb,amsmath}

\usepackage{epsfig}

\usepackage{graphicx}
\usepackage{color}

\newcommand{\emmanuel}[1]{{ #1}}
\newcommand{\vicente}[1]{{ #1}}

\journalname{Granular Matter}

\begin{document}
\title{Dissipative homogeneous Maxwell mixtures: ordering transition in the tracer
limit}

\titlerunning{Ordering transition in dissipative Maxwell mixtures}

\author{Vicente Garz\'o \and Emmanuel Trizac     }


\institute{           Vicente Garz\'o \at
              Departamento de F\'{\i}sica, Universidad
de Extremadura, E-06071 Badajoz, Spain \\
              \email{vicenteg@unex.es}
	      \and
Emmanuel Trizac \at
              Laboratoire de Physique
Th\'eorique et Mod\`eles Statistiques (CNRS UMR 8626), B\^atiment 100,
Universit\'e Paris-Sud, 91405 Orsay cedex, France\\
\email{trizac@lptms.u-psud.fr} }

\date{Received: date / Accepted: date}

\maketitle

\begin{abstract}
The homogeneous Boltzmann equation for inelastic Maxwell mixtures is considered to
study the dynamics of tracer particles or impurities (solvent) immersed in a uniform granular gas (solute).
The analysis is based on exact results derived for a granular binary mixture in the homogeneous cooling state
(HCS) that apply for arbitrary values of the parameters of the mixture (particle masses $m_i$, mole fractions $c_i$,
and coefficients of restitution $\alpha_{ij}$). In the tracer limit ($c_1\to 0$), it is shown that the HCS supports
two distinct phases that are evidenced by the corresponding value of $E_1/E$, the
relative contribution of the tracer species to the total energy.
Defining the mass ratio $\mu\equiv m_1/m_2$,
there indeed exist two critical values $\mu_\text{HCS}^{(-)}$ and $\mu_\text{HCS}^{(+)}$
(which depend on the coefficients of restitution), such that
$E_1/E=0$ for $\mu_\text{HCS}^{(-)}<\mu<\mu_\text{HCS}^{(+)}$ (disordered or normal phase), while
$E_1/E\neq 0$ for $\mu<\mu_\text{HCS}^{(-)}$ and/or $\mu>\mu_\text{HCS}^{(+)}$ (ordered phase).


\end{abstract}

\maketitle

Granular assemblies depart from molecular systems not only from the difference of the length scales
involved, but more importantly in that the interactions among constituents are dissipative \cite{IG03}.
In conjunction with the use of powerful experimental and numerical techniques, the application of non-equilibrium statistical mechanics
to the field has yielded much progress in the last 20 years, whereas the questions were hitherto
more centred on civil and mechanical engineering issues. Among the factors that explain
this upsurge of interest for fundamental approaches, the pioneering work by I. Goldhirsch and
G. Zanetti \cite{GZ93}, pertaining to the clustering instability in an unforced granular gas, should certainly
be recognised. Isaac Goldhirsch subsequently became a leading figure in the granular matter
community, much contributing to the improvement of methods and understanding of models.
As a tribute to his achievements and insights, we address here the particularly simple situation
of a mixture of grains, where spatial homogeneity is enforced, thereby discarding instabilities in
the vein of the clustering phenomenon, but where non trivial out-of-equilibrium phase transitions
take place.

We consider a binary mixture of inelastic Maxwell gases at low density in the homogeneous cooling state (HCS).
The corresponding set of coupled Boltzmann equations for the velocity distributions $f_i({\bf v},t) (i=1,2)$ then read
\begin{equation}
\label{n1} \frac{\partial}{\partial t}f_i=\sum_j\,J_{ij}[{\bf v}|f_i,f_j],
\end{equation}
where the Boltzmann collision operator $J_{ij}[f_i,f_j]$ for dissipative Maxwell mixtures is
\begin{eqnarray}
&& J_{ij}\left[{\bf v}_{1}|f_{i},f_{j}\right] =\frac{\omega_{ij}}{n_j\Omega_d}
\int d{\bf v}_{2}\int d\widehat{\boldsymbol {\sigma }}\left[ \alpha_{ij}^{-1}f_{i}({\bf v}_{1}')f_{j}(
{\bf v}_{2}')\right.\nonumber\\
& & \left.-f_{i}({\bf v}_{1})f_{j}({\bf v}_{2})\right]
\;.
\label{n2}
\end{eqnarray}
Here, $n_i$ is the number density of species $i$, $\Omega_d$ is the total solid angle in $d$ dimensions,
and $\alpha_{ij}\leq 1$ denotes the
(constant) coefficient of restitution  for collisions between particles of species $i$
with $j$. Moreover, ${\bf v}_{1}'={\bf v}_{1}-\mu_{21}\left( 1+\alpha_{12}
^{-1}\right)(\widehat{\boldsymbol {\sigma}}\cdot {\bf g})\widehat{\boldsymbol
{\sigma}}$, ${\bf v}_{2}'={\bf v}_{2}+\mu_{12}\left(
1+\alpha_{21}^{-1}\right) (\widehat{\boldsymbol {\sigma}}\cdot {\bf
g})\widehat{\boldsymbol{\sigma}}$,
where ${\bf g}={\bf v}_1-{\bf v}_2$, $\widehat{\boldsymbol {\sigma}}$ is a unit vector directed along the centers
of the two colliding spheres, and $\mu_{ij}=m_i/(m_i+m_j)$.

The effective collision frequencies $\omega_{ij}$ for collisions
$i$-$j$ are independent of the relative velocities of the colliding particles but can depend on space and time through
its dependence on densities $n_i$ and granular temperature $T$ (see e.g. Ref. \cite{IG08} for a discussion
of this kinetic notion). They can be also seen as free parameters of the model.
In previous works on multicomponent granular systems \cite{G03,GA05,GT10}, $\omega_{rs}$ was chosen to guarantee
that the cooling rate for inelastic Maxwell models (IMM) be the same as that of inelastic hard spheres (IHS).
With this choice, the collision rates $\omega_{ij}$ are (intricate) functions of the temperature ratio $T_1/T_2$,
which precludes analytical progress.
Here, since our problem involves a delicate tracer limit,
we aim at the simplest possible approach. Specifically, we assume that $\omega_{ij}$ is independent of the partial
temperatures $T_i$ of each species but depend on the global temperature $T=c_1T_1+c_2T_2$, $c_i=n_i/(n_1+n_2)$
being the mole fraction of species $i$. Thus, one considers the simple
``plain vanilla'' Maxwell model defined as
$\omega_{ij} = \nu c_j$, where $\nu=An\sqrt{T}$ is an effective collision frequency and the value of the constant $A$
is irrelevant for our purposes.
The form of $\omega_{ij}$ is closer to the original model of Maxwell molecules for ordinary gas mixtures \cite{Maxwell}.
The plain vanilla Maxwell model has been previously considered by several authors \cite{NK02,MP02,MP02a,vanilla} in
homogeneous problems pertaining to granular mixtures.

In the absence of any external energy input, the granular temperature $T(t)$ monotonically decays in time due to
the inelastic nature of the collisions. We are here mainly interested in the time evolution of the partial
temperatures $T_i$ (or equivalently, the partial pressures $p_i=n_iT_i$). In the hydrodynamic regime (for times
much longer than the effective mean free time $\nu^{-1}$), it is expected that all the time dependence of $p_i$
is only through its dependence on the global temperature $T(t)$ \cite{GD99}. The time evolution of $T(t)$ is simply
$\partial_tT=-\zeta T$
where
\begin{equation}
\label{zeta}
\zeta=-\frac{1}{dnT}\sum_{i,j}m_i\int d{\bf v}v^{2}J_{ij}[f_{i},f_{j}]
\end{equation}
is the total cooling rate. In order to solve
the temperature equation,
it is convenient to change to
a new time variable defined as $\tau=\int_{0}^t\; \nu(T(t'))dt'$
yielding $T(t)=T(0)\text{exp}(-\zeta^*\tau)$ where $\zeta^*=\zeta/\nu$. To find the relation between
the ``internal'' clock (related to the average number of collisions suffered per particle) and the ``external''
time $t$, one integrates the relation for $d\tau$ using $\nu\sim \sqrt{T}$ and gets the usual Haff's law \cite{H83}
\begin{equation}
\label{3.2.2}
T(t)=\frac{T(0)}{\left[1+\frac{1}{2}\zeta(0)t\right]^2}.
\end{equation}

The partial pressures $p_i$ can be determined by multiplying both sides of Eq.\ (\ref{n1}) by $ m_i v^2$
and integrating over velocity. Taking into account previous results \cite{G03} derived for dissipative
Maxwell mixtures, one obtains
\begin{equation}
\label{3.3}
\left(\vicente{\frac{\partial}{\partial \tau}}+{\cal L}\right){\cal P}=0,
\end{equation}
where ${\cal P}$ is the column matrix
\begin{equation}
\label{3.4}
{\cal P}=\left(
\begin{array}{c}
p_{1}^*\\
p_{2}^*
\end{array}
\right)
\end{equation}
and ${\cal L}$ is the square matrix
\begin{equation}
\label{3.5}
{\cal L}=\left(
\begin{array}{cc}
\lambda+A_{11}&A_{12}\\
A_{21}&\lambda+A_{22}
\end{array}
\right).
\end{equation}
Here, $p_{i}^*=p_{i}/nT$ and we have introduced the dimensionless quantities
\begin{equation}
\label{3.6}
A_{11}=\frac{\omega_{11}^*}{2\vicente{d}}(1-\alpha_{11}^2)+\frac{2\omega_{12}^*}{\vicente{d}}
\mu_{21}(1+\alpha_{12})
\left[1-\frac{\mu_{21}}{2}(1+\alpha_{12})\right],
\end{equation}
\begin{equation}
\label{3.7}
A_{12}=-\frac{\omega_{12}^*}{\vicente{d}}\frac{\rho_1}{\rho_2}\mu_{21}^2(1+\alpha_{12})^2,
\end{equation}
where $\omega_{ij}^*=\omega_{ij}/\nu$ and $\rho_i=n_im_i$ is the mass density of species $i$. The coefficients $A_{22}$ and $A_{21}$ can be easily obtained from Eqs.\ (\ref{3.6}) and (\ref{3.7}) by change of indices $1\leftrightarrow 2$.
In addition, as pointed out earlier, the temperature $T(t)$ behaves for long times as
$T(t)=T(0)e^{\lambda \tau}$
where
$\lambda$ is a nonlinear function of $\alpha_{ij}$ and the
parameters of the mixture.

\vicente{After a certain kinetic regime lasting a few collision times, one expects that the reduced partial pressures $p_1^*$ and $p_2^*=1-p_1^*$ reach well-defined steady values $p_{1,\text{s}}^*$ and $p_{2,\text{s}}^*$, respectively. These steady values are obtained by solving the homogeneous equation ${\cal L}{\cal P}=0$. This equation} has a nontrivial solution if $\det {\cal L}=0$. This is a
second-degree polynomial equation whose largest root $\lambda_{\text{max}}=\text{max}(\lambda_1,\lambda_2)
=-\zeta^*$ governs the time evolution of the temperature in the long-time limit.
Here, $\lambda_1$ and $\lambda_2$ are the solutions of the equation $\det {\cal L}=0$,
\begin{equation}
\label{3.10}
\lambda_{1,2}=\frac{-(A_{11}+A_{22})\mp\sqrt{(A_{11}-A_{22})^2\vicente{+}4A_{12}A_{21}}}{2d}.
\end{equation}
The steady solution $p_{1,\text{s}}^*$ is given by
\begin{equation}
\label{3.10.1}
p_{1,\text{s}}^*(\lambda)=\frac{A_{12}}{A_{12}-A_{11}-\vicente{\lambda}},
\end{equation}
where $\lambda=\lambda_{\text{max}}$. Consequently, for long times, the time dependence of the
partial pressure $p_1^*(t)$ (or equivalently, the energy ratio $E_1/E\equiv p_1^*$) can be written as
\begin{equation}
\label{3.13}
p_1^*(t)=\frac{A p_{1,\text{s}}^*(\lambda_2)+B p_{1,\text{s}}^*(\lambda_1) e^{-(\lambda_2
-\lambda_1)\nu \tau}}{A+B e^{-(\lambda_2-\lambda_1)\nu \tau}},
\end{equation}
where $A$ and $B$ are constants depending on the initial conditions and the function $p_{1,\text{s}}^*(\lambda)$
is defined in Eq.\ (\ref{3.10.1}). In conclusion, after a relaxation
time of the order of $|(\lambda_2-\lambda_1)\nu|^{-1}$, the energy ratio $p_1^*$
reaches the steady state value $p_{1,\text{s}}^*(\lambda_{\text{max}})$.
As long as the mole fraction $c_1\neq 0$,
one has
$\lambda_1\neq \lambda_2$ for any value of the mass ratio $\mu\equiv m_1/m_2$ and the coefficients of restitution. It must be remarked that the results derived so far coincide with those previously obtained in the one-dimensional case \cite{MP02}.


\begin{figure}
\includegraphics[width=0.85 \columnwidth,angle=0]{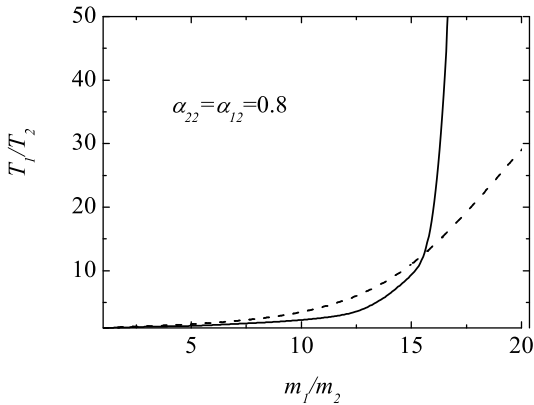}
\caption{Plot of the temperature ratio $T_1/T_2$ \emph{versus} the mass ratio $m_1/m_2$ for $d=3$ in the
case $\alpha_{22}=\alpha_{12}=0.8$.
The solid line is the result for IMM (Inelastic Maxwell Model) while the dashed line is for IHS
(Inelastic Hard Spheres).
\label{fig1}}
\end{figure}
Let us consider now the tracer limit ($c_1\to 0$). In this limit, according to Eq.\ \eqref{3.10},
$\lambda_1$ and $\lambda_2$ become simply
\begin{equation}
\label{3.11}
\lambda_1\to \lambda_1^{(0)}=-\frac{2}{d}\mu_{21}(1+\alpha_{12})\left[1-\frac{\mu_{21}}{2}(1+\alpha_{12})\right],
\end{equation}
\begin{equation}
\label{3.12}
\lambda_2\to \lambda_2^{(0)}=-\frac{1-\alpha_{22}^2}{2d}.
\end{equation}
The root $\lambda_1^{(0)}$ gives the time behavior of the temperature $T_1$ of the tracer particles while
$\lambda_2^{(0)}$ is associated with the time evolution of the granular temperature $T_2$ of the excess component.
For given values of the coefficients of restitution $\alpha_{22}$ and $\alpha_{12}$, it can be easily proved that
$\lambda_2^{(0)}>\lambda_1^{(0)}$ if the mass ratio $\mu$ lies in the range
$\mu_{\text{HCS}}^{(-)}<\mu<\mu_{\text{HCS}}^{(+)}$, where the critical mass ratios $\mu_{\text{HCS}}^{(-)}$ and
$\mu_{\text{HCS}}^{(+)}$ are obtained from the condition $\lambda_2^{(0)}=\lambda_1^{(0)}$. They are given by
\begin{equation}
\label{3.15other}
\mu_{\text{HCS}}^{(-)}=\frac{\alpha_{12}-\sqrt{\frac{1+\alpha_{22}^2}{2}}}
{1+\sqrt{\frac{1+\alpha_{22}^2}{2}}}, \quad \mu_{\text{HCS}}^{(+)}=\frac{\alpha_{12}+\sqrt{\frac{1+\alpha_{22}^2}{2}}}
{1-\sqrt{\frac{1+\alpha_{22}^2}{2}}}.
\end{equation}
On the other hand, if the mass ratio $\mu$ is smaller (resp. larger) than $\mu_{\text{HCS}}^{(-)}$
(resp. $\mu_{\text{HCS}}^{(+)})$, then $\lambda_1^{(0)}>\lambda_2^{(0)}$. For elastic collisions ($\alpha_{22}=\alpha_{12}=1$), $\mu_{\text{HCS}}^{(-)}=0$ and $\mu_{\text{HCS}}^{(+)}\to \infty$ and so $\lambda_2^{(0)}$ is always larger than $\lambda_1^{(0)}$.

The above results show clearly that there are two different regimes of behavior. When $\lambda_2^{(0)}>\lambda_1^{(0)}$,
the tracer temperature $T_1(t)$ is enslaved to the granular temperature $T_2(t)$ and so, the temperature ratio
asymptotically reaches the steady state value $p_{1,\text{s}}^*(\lambda_2^{(0)})/c_1$. More explicitly,
\begin{equation}
\label{3.14}
\lim_{t\to \infty}\frac{T_1(t)}{T_2(t)}=\frac{2\mu_{12}\mu_{21}(1+\alpha_{12})^2}{4\mu_{21}(1+\alpha_{12})
\left[1-\frac{\mu_{21}}{2}(1+\alpha_{12})\right]-1+\alpha_{22}^2}.
\end{equation}
On the other hand, if $\lambda_1^{(0)}>\lambda_2^{(0)}$, then the combination $A_{12}-A_{11}-\lambda_1^{(0)}$
vanishes in the tracer limit so that, according to Eq.\ (\ref{3.10.1}), the temperature ratio
$p_{1,\text{s}}^*(\lambda_1^{(0)})/c_1$ tends to infinity. This latter case corresponds to an extreme breakdown of
the energy equipartition since the tracer particles are very energetic compared with the gas particles and the
impurities essentially scatter off a static fluid background. The transition toward the heavy-impurity phase (i.e.,
when $\mu>\mu_{\text{HCS}}^{(+)}$) was already found by Ben-Naim and Krapivksy \cite{NK02} in their analysis on the
velocity statistics of an impurity immersed in a uniform granular fluid. The light-impurity phase (i.e.,
when $\mu<\mu_{\text{HCS}}^{(-)}$) is not reported in Ref.\ \cite{NK02}; it only appears when $\alpha_{12}>\sqrt{(1+\alpha_{22}^2)/2}$. Thus, this new phase
disappears (since $\mu_{\text{HCS}}^{(-)}$ becomes negative) when $\alpha_{12}=\alpha_{22}$ or when
$\alpha_{12}<1/\sqrt{2}$. It must also be remarked that a similar non-equilibrium transition has been found for IHS \cite{SD01} showing that, in the anomalous or ordered phase, the ratio of the mean square velocities for the impurity and fluid particles $T_1m_2/T_2m_1$ is finite (and so, the temperature ratio is infinite) even for extremely large mass ratios ($m_1/m_2\to \infty$). Although the transition to the heavy-impurity phase detected for IMM occurs in general for large mass ratios (for instance, $\mu_{\text{HCS}}^{(+)}\simeq 18.05$ for $\alpha_{22}=\alpha_{12}=0.8$),
the transition phenomenon found in Ref.\ \cite{SD01} is less pronounced for hard spheres interaction since at a practical point one needs to consider much bigger values of the mass ratio for IHS to find the above transition.

The expression (\ref{3.14}) for the temperature ratio derived in the tracer limit ($c_1\to 0$) agrees with the one obtained in Ref.\ \cite{NK02} from the Boltzmann-Lorentz equation.
It appears that in general $T_1/T_2\neq 1$, although
Eq.\ \eqref{3.14} shows that energy equipartition occurs when the mass
ratio is given by
\emmanuel{$\mu_\text{eq}=(1+\alpha_{22}^2-2\alpha_{12}^2)/(1-\alpha_{22}^2)$,
provided $\alpha_{22}\neq 1$}.
Moreover, when the particles
of the gas collide elastically ($\alpha_{22}=1$), then $T_1/T_2=(1+\alpha_{12})/[2+(1-\alpha_{12})(\mu_{21}/\mu_{12})]$.
This expression coincides with the one obtained \cite{MP99} for IHS. Beyond this case, the dependence of $T_1/T_2$ on
the parameters of the system in the normal phase is different from that of hard spheres \cite{GD99}. Figure \ref{fig1}
shows the dependence of the temperature ratio on the mass ratio for $\alpha_{22}=\alpha_{12}=0.8$. In this case,
$\mu_{\text{HCS}}^{(-)}\simeq -0.055$ and $\mu_{\text{HCS}}^{(+)}\simeq 18.05$ and so, there is only heavy-impurity phase.
The $\mu$-dependence of the temperature ratio for IHS (with the same diameter for the tracer and gas particles) is also
shown for comparison. We observe that IMM capture well the trends of IHS, except of course close to the mass critical value
where $T_1/T_2$ grows very fast with $m_1/m_2$ for IMM. This growing is less dramatic for IHS.
Likewise, the plain vanilla approach
exaggerates the features of the more refined Maxwell model alluded to above \cite{G03,GA05,GT10};
the trends evidenced, though, appear to be robust.

\begin{figure}
\includegraphics[width=0.85 \columnwidth,angle=0]{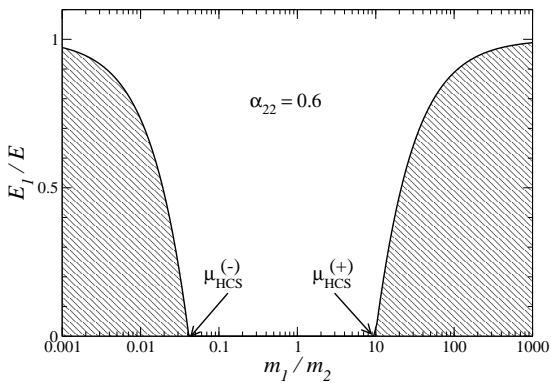}
\caption{Plot of the order parameter $E_1/E$ versus the mass ratio $m_1/m_2$ for $\alpha_{12}=0.9$ and
$\alpha_{12}=0.6$. The hatched regions indicate the ordered phases while the arrows correspond to the critical mass ratios $\mu_{\text{HCS}}^{(-)}\backsimeq 0.041$ and $\mu_{\text{HCS}}^{(+)}\backsimeq 9.83$. \label{fig2}}
\end{figure}

We now explore the physical consequences of the existence of critical mass ratios in the tracer limit.
In order to analyze this point,
we consider for instance the energy ratio $E_1/E\equiv p_{1,\text{s}}^*$. For $c_1\neq 0$, the expression of $E_1/E$
is given by Eq.\ (\ref{3.10.1}) where the explicit dependence of $A_{11}$ and $A_{12}$ on $c_1$, $\mu$ and $\alpha_{ij}$
are given by Eqs.\ (\ref{3.6}) and (\ref{3.7}), respectively. If $c_1\to 0$, Eq.\ (\ref{3.10.1}) becomes
\begin{equation}
\label{3.15}
\frac{E_1}{E}\approx c_1 \frac{A_{12}^{(1)}}{(A_{12}^{(1)}-A_{11}^{(1)})c_1-A_{11}^{(0)}-\vicente{\lambda}},
\end{equation}
where
\begin{equation}
\label{3.16}
A_{11}^{(0)}=-\lambda_1^{(0)}=\frac{2}{\vicente{d}}\mu_{21}(1+\alpha_{12})
\left[1-\frac{\mu_{21}}{2}(1+\alpha_{12})\right],
\end{equation}
\begin{equation}
\label{3.17}
A_{11}^{(1)}=\frac{1-\alpha_{11}^2}{2\vicente{d}}, \quad
A_{12}^{(1)}=-\frac{\mu_{21}}{\vicente{d}}\mu_{12}(1+\alpha_{12})^2.
\end{equation}
Equation (\ref{3.15}) holds for $\lambda_1$ and $\lambda_2$ which are still functions of $c_1$. To first order in $c_1$,
they can be written as
\begin{equation}
\label{3.18} \lambda_1(c_1)\approx \lambda_1^{(0)}+\lambda_1^{(1)} c_1, \quad
\lambda_2(c_1)\approx \lambda_2^{(0)}+\lambda_1^{(1)} c_1,
\end{equation}
where $\lambda_1^{(0)}$ and $\lambda_2^{(0)}$ are given by Eqs.\ (\ref{3.11}) and (\ref{3.12}), respectively.
The expressions of $\lambda_1^{(1)}$ and $\lambda_2^{(1)}$ can be obtained from their forms (\ref{3.10})
and (\ref{3.10.1}) for arbitrary $c_1$.
After some algebra, one gets
\begin{equation}
\label{3.19}
\lambda_1^{(1)}=-\frac{1-\alpha_{11}^2}{2d}-\frac{\mu_{21}^2\mu_{12}^2
(1+\alpha_{12})^4}
{d^2(\lambda_2^{(0)}-\lambda_1^{(0)})},
\end{equation}
\begin{eqnarray}
\label{4.18}
\lambda_2^{(1)}&=&-\frac{2}{d}\mu_{12}(1+\alpha_{12})
\left[1-\frac{\mu_{12}}{2}(1+\alpha_{12})\right]\nonumber\\
& & -\frac{\mu_{21}^2\mu_{12}^2
(1+\alpha_{12})^4}
{d^2(\lambda_1^{(0)}-\lambda_2^{(0)})}.
\end{eqnarray}
It must be noted that if $\lambda=\lambda_2^{(0)}$ in Eq.\ (\ref{3.15}),
then [according to Eqs.\ (\ref{3.12}) and
(\ref{3.16})] $A_{11}^{(0)}+\vicente{\lambda_2^{(0)}}\neq 0$ and
so the energy ratio $E_1/E=0$ when $c_1\to 0$ as expected. However, if
$\lambda=\lambda_1^{(0)}$ in Eq.\ (\ref{3.15}), $A_{11}^{(0)}+\vicente{\lambda_1^{(0)}}=0$ and so, $E_1/E\neq 0$.
More specifically, by taking the tracer limit in Eq.\ (\ref{3.15}) when $\lambda=\lambda_1$ one gets
\begin{eqnarray}
\label{3.20} & & \lim_{c_1\to
0}\frac{E_1}{E}=\frac{A_{12}^{(1)}}{A_{12}^{(1)}-A_{11}^{(1)}-\vicente{\lambda_1^{(1)}}}\nonumber\\
&=&\frac{\alpha_{22}^2-1+4\mu_{21}(1+\alpha_{12})\left[1-\frac{\mu_{21}}{2}(1+\alpha_{12})\right]}
{\alpha_{22}^2-1+2\mu_{21}(1-\alpha_{22}^2)}.
\end{eqnarray}
Note that although $A_{11}^{(1)}$ and $\lambda_1^{(1)}$ depend on the coefficient of restitution $\alpha_{11}$,
the energy ratio is independent on collisions among tracer particles themselves.
This means that one could neglect the Boltzmann collision operator $J_{11}[f_1,f_1]$ in the kinetic equation of
the one-particle velocity distribution function $f_1$ (Boltzmann-Lorentz description). This is
quite a natural
assumption when one analyzes the tracer problem. In addition, it is also usual to assume that the presence of
tracer particles does not affect the state of the solvent (excess component) and so, collisions of type 2-1 can
be neglected (closed Boltzmann equation for the gas). On the other hand, in the heavy or light impurity phase,
one needs to consider the contributions coming from the Boltzmann operator $J_{21}[f_2,f_1]$ to get Eq.\ (\ref{3.20}).
This is clearly shown in the Appendix \ref{appA} where the expression (\ref{3.20}) is derived from an alternative route.

\begin{figure}
\includegraphics[width=0.85 \columnwidth,angle=0]{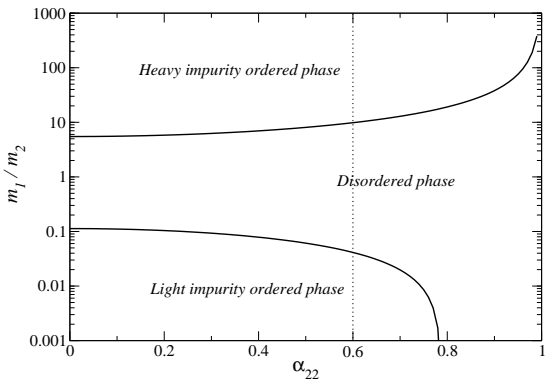}
\caption{Phase diagram in the dissipation/mass ratio plane, for a fixed value of
tracer-fluid inelasticity $\alpha_{12}=0.9$. The vertical dotted line indicates the
cut corresponding to Fig.\ \ref{fig2}.
\label{fig3}}
\end{figure}

In conclusion, when $\mu_{\text{HCS}}^{(-)}<\mu<\mu_{\text{HCS}}^{(+)}$, the temperature ratio $T_1/T_2$ is finite and so, the energy ratio $E_1/E=0$. On the other hand, if $\mu<\mu_{\text{HCS}}^{(-)}$ or $\mu>\mu_{\text{HCS}}^{(+)}$, the temperature ratio diverges to
infinity and the energy ratio becomes {\em finite}. This change of behavior is similar to an ordering process where the impurity is enslaved to the host fluid ($E_1/E=0$), or carries a finite fraction of the total kinetic energy of the system ($E_1/E\neq 0$). The latter situation can be referred to as the ``ordered'' phase (extreme breakdown of the energy equipartition) while the first one can be coined ``disordered'' phase. Figures \ref{fig2} and \ref{fig3} illustrate the transition phenomenon found here. Specifically, Fig.\ \ref{fig2} shows the energy ratio $E_1/E$ as a function of the mass ratio $m_1/m_2$ for $\alpha_{22}=0.6$ and $\alpha_{12}=0.9$. In this case, according to Eq.\ \eqref{3.15other}, one gets $\mu_{\text{HCS}}^{(-)}\backsimeq 0.041$ and $\mu_{\text{HCS}}^{(+)}\backsimeq 9.83$. It is apparent that, for asymptotically small or large mass ratios, the contribution of the impurities to the total energy can be even larger than that of the host gas. Figure \ref{fig3} shows a phase diagram in the $\{\alpha_{22}, m_1/m_2\}$-plane at $\alpha_{12}=0.9$. The light impurity ordered phase appears for values of the mass ratio $\mu \lesssim 0.113$ provided that $\alpha_{22}\lesssim 0.787$ while the heavy impurity ordered phase is present for mass ratios $\mu \gtrsim 5.487$ in the complete range of values of the coefficient of restitution $\alpha_{22}$. Finally, it must be noted that a similar non-equilibrium phase transition has been found for a sheared granular mixture \cite{GT11}.

\appendix
\section{Energy ratio in the ordered phase}
\label{appA}

In this Appendix  we will obtain the expression (\ref{3.20}) of the energy ratio $E_1/E$ in the \emph{ordered}
phase from the condition $\zeta_1=\zeta_2$ (this is the condition to determine the temperature ratio in the
HCS \cite{GD99}) when the collisions among tracer particles themselves are neglected. Here, $\zeta_i$ refers
to the partial cooling rate associated to species $i$. In the case of IMM, the cooling rates $\zeta_i$ have
been exactly obtained in Ref.\ \cite{G03}. In the tracer limit ($c_1\to 0$), if one neglects the effect
of collisions 1-1, one has $\zeta_1\simeq \zeta_{12}$ and 
$\zeta_2=\zeta_{21}+\zeta_{22}$ where
\begin{equation}
\label{a1}
\zeta_{12}=-\lambda_1^{(0)}\nu=
\frac{2}{d}\nu\mu_{21}(1+\alpha_{12})\left[1-\frac{\mu_{21}}{2}(1+\alpha_{12})\right],
\end{equation}
\begin{equation}
\label{a2}
\zeta_{22}=-\lambda_2^{(0)}\nu=\frac{1-\alpha_{22}^2}{2d}\nu,
\end{equation}
\begin{equation}
\label{a3}
\zeta_{21}=-\frac{\mu_{21}\mu_{12}}{d}\nu(1+\alpha_{12})^2\frac{p_{1,\text{s}}^*}{1-p_{1,\text{s}}^*},
\end{equation}
where use has been made of the identity $p_{2,\text{s}}^*=1-p_{1,\text{s}}^*$. Moreover, we are considering
the ordered phase and so $p_{1,\text{s}}^*$ is finite (and $T_1/T$ is infinite). If one neglects the tracer
collisions, the condition to get $p_{1,\text{s}}^*$ reduces to $\zeta_{12}=\zeta_{22}+\zeta_{21}$
Substitution of Eqs.\ (\ref{a1}), (\ref{a2}), and (\ref{a3}) into the previous relation
yields
\begin{equation}
\label{a5}
\frac{d}{2}(\lambda_1^{(0)}-\lambda_2^{(0)})=\frac{1}{2}\mu_{21}\mu_{12}(1+\alpha_{12})^2
\frac{p_{1,\text{s}}^*}{1-p_{1,\text{s}}^*}.
\end{equation}
The solution to Eq.\ (\ref{a5}) is
\begin{equation}
\label{a8}
p_{1,\text{s}}^*=\frac{\alpha_{22}^2-1+4\mu_{21}(1+\alpha_{12})\left[1-\frac{\mu_{21}}{2}(1+\alpha_{12})
\right]}
{\alpha_{22}^2-1+2\mu_{21}(1-\alpha_{12}^2)}.
\end{equation}
Upon writing this equation the explicit forms of $\lambda_1^{(0)}$ and $\lambda_2^{(0)}$ have been considered.
The expression (\ref{a8}) coincides with Eq.\ (\ref{3.20}).

\begin{acknowledgements}

The research of V.G. has been supported by
the Ministerio de Ciencia e Innovaci\'on  (Spain) through grant No. FIS2010-16587,
partially financed by FEDER funds and by the Junta de Extremadura (Spain) through Grant No. GRU10158.

\end{acknowledgements}

\end{document}